\documentstyle[12pt,epsf,bezier,axodraw]{article}
\begin{document}
\setlength{\baselineskip}{0.33 in}
%%%%%%%%%%%%%%%
\catcode`@=11
% Redefine caption to put text and formulas in smaller font
\long\def\@caption#1[#2]#3{\par\addcontentsline{\csname
  ext@#1\endcsname}{#1}{\protect\numberline{\csname
  the#1\endcsname}{\ignorespaces #2}}\begingroup
    \small
    \@parboxrestore
    \@makecaption{\csname fnum@#1\endcsname}{\ignorespaces #3}\par
  \endgroup}
\catcode`@=12
%%%%%%%%%%%%%%%%%%%%%%%%%%%%%%%%%%%%%%%%%%%%%%%%%%%%%%%%%%%%
\newcommand{\newc}{\newcommand}
\newc{\gsim}{\lower.7ex\hbox{$\;\stackrel{\textstyle>}{\sim}\;$}}
\newc{\lsim}{\lower.7ex\hbox{$\;\stackrel{\textstyle<}{\sim}\;$}}
%%%%%%%%%%%%%%%%%% Reference Defs %%%%%%%%%%%%%%%%%
\def\NPB#1#2#3{Nucl. Phys. {\bf B#1} #3 (19#2)}
\def\PLB#1#2#3{Phys. Lett. {\bf B#1} #3 (19#2)}
\def\PRD#1#2#3{Phys. Rev. {\bf D#1} #3 (19#2)}
\def\PRB#1#2#3{Phys. Rev. {\bf B#1} #3 (19#2)}
\def\PRL#1#2#3{Phys. Rev. Lett. {\bf#1} #3 (19#2)}
\def\PRT#1#2#3{Phys. Rep. {\bf#1} #3 (19#2)}
\def\MODA#1#2#3{Mod. Phys. Lett. {\bf A#1} #3 (19#2) }
\def\ZPC#1#2#3{Zeit. f\"ur Physik {\bf C#1} #3 (19#2) }
\def\ZPA#1#2#3{Zeit. f\"ur Physik {\bf A#1} #3 (19#2) }
%%%%%%%%%%%%%%%%%%%%%%%%%%%%%%%%%%%%%%%%%%%%%%%%%%%%%%%%%%%%%
\def\bdm{\begin{equation}}
\def\edm{\end{equation}}
\def\bea{\begin{eqnarray}}
\def\eea{\end{eqnarray}}
\def\sp{symmetric phase }
\def\bsp{broken symmetry phase }
\def\cc{coupling constant }
\def\bm2{\mu_0^2}
\def\rc{renormalization condition }
\def\rcz{renormalization constant }
\def\ET{Equivalence Theorem }
\def\ssb{Spontaneous Symmetry Breaking }
\def\lpgvb{longitudinaly polarized gauge vector bosons }
\def\GB{Goldstone bosons }
\def\d0{\delta_0}
\def\GF{Green functions }
\def\WI{Ward identities }
\def\tl{tree-level }
\def\1l{one-loop }
\def\rcs{renormalization conditions }
\def\rczs{renormalization constants }
\def\M{{\cal M}}
\def\chis{\chi^{\star}}
\def\km{k^{\mu}}
\def\akl{a(k,\lambda_{\ell}}
\def\idelt{i(2\pi)^4\delta^4(\cdots)}
\def\lamlong{\lambda_{\ell}}
\def\tf{\tau-function}
\def\tsf{\tau^{\star}-function}
\def\phis{\phi^{\star}}
\def\ra{\rightarrow}
\def\g2{\frac{g^2}{16 \pi^2}}
\def\fac{\g2 \frac{m^2}{M^2}}
\def\zz{\ZigZag}
\def\dl{\DashLine}
\def\dcarc{\DashCArc}
\def\De{\Delta_{\epsilon}}
\def\sm{Standard Model }
\def\lra{\leftrightarrow}
\def\oa{\stackrel{\leftrightarrow}}
\def\SMEs{S-Matrix elements }
%%%%%%%%%%%%%%%%%%%%%%%%%%%%%%%%%%%%%%%%%%%%%%%%%%%%%%%%
\vsize 8.7in
\def\singlespace{\baselineskip 11.38 pt}
\def\halfagainspace{\baselineskip 17.07 pt}
\def\doublespace{\baselineskip 22.76 pt}
\def\medspace{\baselineskip 17.07 pt}
\font\headings=cmbx10 scaled 1200
\font\title=cmbx10 scaled 1200
%halfagainspace
\singlespace

\begin{titlepage}
\begin{flushright}
{\large
hep-ph/9804392\\
Aug 31, 1998 \\
}
\end{flushright}
\vskip 2cm
\begin{center}
{\Large {\bf Unstable Higgs Particle and the Equivalence Theorem}}

\vskip 1cm
{\Large 
R.S. Willey\footnote{E-mail: {\tt willey@vms.cis.pitt.edu}}$^{a}$\\}
\vskip 2pt
\smallskip
$^{a}${\large\it Department of Physics and Astronomy\\
 University of Pittsburgh, Pittsburgh, PA 15260, USA}\\
\end{center}

%%%%%%%%%%%%%%%%%%%%%%%%%%%%%%%%%
\vspace*{.3in}
\begin{abstract}
We consider the subtleties involved in the application of the \ET to  
the decay of an unstable Higgs particle.
This is formally justified from consideration of unitarity and resonant elastic scattering of the stable decay prodiucts.  
By explicit perturbative calculation we find that the imaginary parts            of the one-loop
amplitudes for that decay are in agreement with the \ET only when one
includes the imaginary parts generated by evaluating the tree-level 
amplitudes at the complex on-shell value of the four-momentum squared 
of the unstable particle.

\end{abstract}
\end{titlepage}
\setcounter{footnote}{0}
\setcounter{page}{2}
\setcounter{section}{0}
\setcounter{subsection}{0}
\setcounter{subsubsection}{0}
\setcounter{equation}{0}
%%%%%%%%%%%%%%%%%%%%%%%%%%%%%%%%%%%%%
%%%%%%%%%%%%%%%%%%%%%%%%%%%%%%%%%%%%%

\newpage
\section{\bf Introduction}
The \ET  is a property of gauge theories with \ssb .
The gauge vector bosons 
acquire mass via the vev of a scalar (Higgs) field. In a $R_{\xi}$ gauge 
there are also one or more unphysical wouldbe Goldstone bosons. The \ET 
asserts that, in the limit in which the energies are all much 
greater than the vector boson masses, \SMEs
 with external longitudinal polarized gauge vector bosons
 are equal to (up to constant overall phase)
\SMEs which have the external longitudinal polarized 
vector bosons replaced by external wouldbe Goldstone bosons of the 
same four-momenta. For Green Functions, the statement is true at tree 
level, but loop corrections may introduce renormalization scheme dependent 
correction factors. There is a substantial literature on this subject
which may be traced from the recent review by He,Kuang,Yuan \cite{hky}.

  The derivations all rely, explicitly or implicitly, on the LSZ reduction 
formulas to relate perturbatively computed renormalized $\tf$s (Green 
Functions) to \SMEs. This in turn implies that external lines in Feynman 
diagrams (for \SMEs) represent stable particles in asymptotic states 
(physically, sufficiently long lived to reach detectors outside the interaction
 region). One may then ask if the \ET can be applied to the decay of an
unstable Higgs boson that is known only by the detection of its stable decay 
products. Is the invariant matrix-element for the decay 
of a heavy Higgs particle into two  longitudinal polarized gauge bosons 
the same as that for the decay of the Higgs particle into two wouldbe \GB ?

   The simplest model in which one can study these matters is the Abelian 
Higgs model. In section two we go over the renormalization and BRST 
structure of this model. In section three we determine the finite ratios 
of field strength \rcz required for LSZ to relate renormalized $\tf$s 
to \SMEs when all the particles are stable. In section four we take up 
the questions raised by the instablity of the Higgs boson. We give a 
heueristic continuity argument and then a derivation based on unitarity 
of the resonance contribution to the scattering of stable particles that
formally justifies application of the \ET to this case.
 In section five 
we check the general results by explicit perturbative (tree plus one-loop) 
calculation in the Abelian Higgs model. In section six the corresponding 
perturbative calculations are done for the \sm of the electroweak  
interactions.          

In both cases, we find that the simple 
imaginary parts of the \1l amplitudes (which directly reflect the instablity) 
agree only after
they are combined with the imaginary parts generated by evaluating the 
tree-level amplitudes at the complex on-shell value of the four-mommentum 
squared of the unstable particle.

\section{\bf Abelian Higgs Model and BRST}

 The 'classical' Lagrangian is
\bdm
    {\cal L}_{cl} = -\frac{1}{4}A_{0_{\mu\nu}}A_0^{\mu\nu}+|D_0\phi_0|^2 -
     \mu_0^2|\phi_0|^2 -\lambda_0 (|\phi_0|^2)^2   \label{Lcl}
\edm
where
\bdm
  A_0^{\mu\nu}= \partial^{\mu}A_0^{\nu}-\partial^{\nu}A_0^{\mu},\;\;\;\;
   D_0^{\mu}\phi_0 = (\partial^{\mu}+ig_0A_0^{\mu})\frac{1}{\sqrt{2}}(h_0-
   i\chi_0)    \label{AD}
\edm
The subscript zero denotes canonical unrenormalized fields and bare 
parameters.
 
   Perturbatively, for $\bm2 > 0$, the system is in the symmetric phase 
with $<0|h_0|0> = 0$. For $\bm2 < 0$, the system is in spontaneously broken 
symmetry phase with 
 $$
<0|\phi_0|0> = \frac{1}{\sqrt{2}}<0|h_0|0> = \frac
{1}{\sqrt{2}}V_0.
$$
 we separate $h_0=V_0+\hat{h}_0$.

  We choose a gauge fixing function of the Feynman-t'Hooft class to 
cancel the tree level $\chi -\partial A$ mixing generated in (\ref{Lcl}) by       the above 
field shift in the \bsp .

\bdm    
  {\cal L}_{GF}  = -\frac{1}{2\xi_0} F_0^2   \label{GF}
\edm
\bdm    F_0 = \partial A_0 + \xi_0 g_0 \kappa_0 V_0 \chi_0   \label{F0}
\edm 
The second gauge parameter $\kappa_0$ is required for consistent renormalization 
of $F_0$. (See below ).
With this gauge fixing the Fadeev-Popov effective Lagrangian is
\bea
 {\cal L}_{FP} & = & const\;\bar{c}_0\;\frac{\delta F_0}{\delta \omega}\;c_0 
                                   \nonumber  \\
               & = & \bar{c}_0(-\partial^2 -\xi_0 g_0^2 \kappa_0 V_0 
(V_0+\hat{h}_0))c_0   
                        \label{FP}
\eea

The BRST transformations under which ${\cal L}_{cl}$ and ${\cal L}_{GF} + 
{\cal L}_{FP}$ are invariant are:
\bea
 \d0 A_0^{\mu}= -\partial^{\mu} c_0, & \d0 \hat{h}_0 = g_0\chi_0 c_0 & 
 \d0 \chi_0 = -g_0(V_0+\hat{h}_0)c_0  \nonumber  \\
 \d0 c_0 = 0, & \d0 \bar{c}_0 = -\frac{1}{\xi_0}F_0, & \d0 F_0 = 0 \label{BRST}  
\eea

We now reparametrize the Lagrangian in terms of renormalized fields and 
parameters. We require that this be done in such a way that the 
renormalized fields and parameters satisfy the same BRST relations 
(\ref{BRST}) and Ward identites as the unrenormalized fields and 
parameters.

\bea
 A_0 = \sqrt{Z_A}A, & \;\;(\hat{h}_0,\chi_0,V_0) = \sqrt{Z}(\hat{h},\chi,V) &   
                                               \nonumber  \\
  \bm2 = Z_{\mu^2}\mu^2, & \lambda_0 = Z_{\lambda}\lambda, & g_0=Z_g g 
                                     \nonumber   \\
  \xi_0 = Z_{\xi}\xi,  & \kappa_0 = Z_{\kappa}\kappa  &   \nonumber   \\
  c_0 = \sqrt{Z_c} c, &  \bar{c}_0 = \sqrt{Z_{\bar{c}}}\bar{c}  &  \label{R}
\eea
The vertex renormalization factors are:
\bdm
 Z_{4\lambda}= Z^2 Z_{\lambda},\;\;Z_{3g}=Z\sqrt{Z_A}Z_g,\;\;
  Z_{4g}= Z Z_A Z_g^2   \label{VZ} 
\edm
The Abelian Ward identities provide the relations
\bdm
 Z_{\xi}=Z_A,\;\; Z_{3g}= Z\;\;(Z_g = Z_A^{-\frac{1}{2}}),\;\;Z_{4g}=Z  \label{WI}
\edm

With these definitions and Ward identities, the BRST transformations of 
$A_0,\hat{h}_0,\chi_0$ all renormalize as 
$$\d0 X_0 = Y_0,\;\;\; \d0 X= \sqrt{\frac{Z_c}{Z_A}}Y  $$
So define renormalized BRST variation
\bdm
 \d0 = \sqrt{\frac{Z_c}{Z_A}}\delta     \label{delta}
\edm
The transformation of $c_0$ renormalizes trivially. For the transformation 
relating $\do \bar{c}_0$ to $F_0$ we have
\bdm
 F_0=\partial A_0 +\xi_0 g_0 \kappa_0 V_0 \chi_0 = 
   \sqrt{Z_A}\partial A+ Z_{\xi}Z_g Z_{\kappa}Z \xi g \kappa V\chi \label{F} 
\edm 
By the Ward identities, $Z_{\xi}Z_g = \sqrt{Z_A}$, so $Z_{\kappa}Z$ must be 
finite. Any finite residual can be absorbed in the as yet unspecified 
constant $\kappa$. So we take
\bdm 
 Z_{\kappa} = Z^{-1}  \label{ZK}
\edm
Then
\bdm
 F_0 = \sqrt{Z_A}(\partial A + \xi g \kappa V\chi)=\sqrt{Z_A} F \label{F2}
\edm
The last BRST transformation is then
\bdm
 \d0 \bar{c}_0 = -\sqrt{\frac{Z_c Z_{\bar{c}}}{Z_A}}\delta \bar{c} 
               = -\frac{1}{\sqrt{Z_A}}\frac{F}{\xi}  \label{F3}
\edm
which fixes
\bdm
  Z_c Z_{\bar{c}}=1   \label{Zc}
\edm
With these specifications, the renormalizations (\ref{R}) preserve the 
structure of the BRST transformations (\ref{BRST}). So the renormalized 
\GF satisfy the same \WI and BRST relations as the (dimensionaly 
regularized) unrenormalized \GF. 

    The gauge parameter $\xi$ will be left free, so the analysis will 
be for any $R_{\xi}$ gauge. The second gauge parameter will be fixed by 
the considerations given now. First $\kappa$ must be one at tree-level 
to cancel the tree-level $\chi -\partial A$ mixing. Next is the treatmment 
of the vev V and the vector boson mass M. The exact renormalized vev V 
is determined order by order in perturbation theory, as a function of the 
renormalized parameters of the theory, by the requirement 
\bdm
      <0|\hat{h}|0> = 0         \label{hat}
\edm
that is, all tadpoles vanish. At tree-level,
\bdm
      V^{0-loop}\equiv v = \sqrt{\frac{-\mu^2}{\lambda}},\;\;\;
            (\mu^2 < 0)   \label{v}
\edm
Beyond \tl V depends also on g, and particularly, on the gauge parammeter 
$\xi$. So the vector boson mass M is specified in terms of the gauge 
invariant \tl vev.
\bdm
          M = g v     \label{M}
\edm
So we will take $\kappa$ to have a factor $\frac{v}{V}$. Finally, to make 
the \ET as a relation between S-matrix elements come out with no extra 
constant factors, we will include a factor 
 $\sqrt{{\frac{Z}{Z^{\star}_{\chi}}}}$ 
in $\kappa$
\bdm
   \kappa = \sqrt{\frac{Z}{Z^{\star}_{\chi}}}\frac{v}{V}    \label{K}
\edm 
Here, we distinguish between $Z$, the renormalization of $\chi_0$ introduced 
in (\ref{R}) with no commitment as to the \rc required to fix its value, 
and $Z^{\star}_{\chi}$ which is the renormalization of $\chi_0$ fixed by the  
\rc that the pole of the complete renormalized $\chi$ two-point function
have unit residue
(derivative of the renormalized self-energy function vanishes at $k^2 = 
M_{\chi}^2$). The field
\bdm
   \chi^{\star} = \sqrt{\frac{Z}{Z^{\star}_{\chi}}}\chi    \label{star}
\edm
is the renormalized field which satisfies the LSZ asymptotic condition with 
no finite field renormalization factor. With these specifications, the 
renormalized F is 
\bdm
  F = \partial A + \xi M \chis       \label{F4}
\edm
Among the \rcs necessary to complete the sppecification of the renormalization 
(\ref{R}), we will always include the unit residue condition for the 
physical vector boson.
\bdm
  A_0 = \sqrt{Z^{\star}_A}A,\;\;\; A = A^*   \label{A} 
\edm

The BRST analysis   provides the result \cite{e}
\bdm
     <\beta,phys'|F|\alpha,phys'> = 0     \label{0F}
\edm
Physical states may be taken to be in,out-states of physical particles 
whose asymptotic in,out-fields ($g\rightarrow 0$) are BRST invariant.
Since (\ref{BRST}) has the $\chi$ field BRST invariant in the limit 
$g\ra 0$, just as the $\hat{h}$ field, the states $|phys'>$ in (\ref{0F})
may include wouldbe Goldstone bosons,
 as well as Higgs scalars and vector bosons in a 
physical polarization state (spacelike transverse or longitudinal in three 
dimensions, not timelike).

\section{\bf The Equivalence Theorem}
The \ET is now a simple application of the LSZ reduction formulas, combined 
with(\ref{0F}). We will arrange \rcs so that the masses in the renormalized 
free propagators are the physical masses (modulo problems posed by unstable 
particles). That is, \rczs are chosen so that renormalized self energy 
functions vanish on their respective mass shells.
\bdm
 M_A^2 = g^2 v^2 \equiv M^2, \;\;\; m_h^2=2\lambda v^2 \equiv m^2,\;\;\; 
    M_{\chi}^2 = \xi M^2   \label{masses} 
\edm

We start with the LSZ reduction formula to take a single $\chi-particle$ 
out of an in-state.
\bdm
   <\beta|\alpha,\chi(k)> = i\int\;(dx)e^{-ikx}(\partial^2 +\xi M^2)
     <\beta|\chis(x)|\alpha>  \\
   = i(2\pi)^4\delta^4(p_{\beta}-p_{\alpha}-k)\M_{\chi}(\beta;\alpha,
                                                   \chi(k)) \label{xreduc}
\edm
For a vector boson of polarization $\lambda$
\bea
 <\beta|\alpha,a(k,\lambda)> = -i\int\;(dx)e^{-ikx}\varepsilon^{\mu}(k,\lambda)
  K_{\mu \nu}<\beta|A^{\nu^*}(x)|\alpha>  &  &  \hspace{1.5in} \nonumber  \\
       = i(2 \pi)^4\delta^4(p_{\beta}-p_{\alpha}-k)\varepsilon^{\mu}
         \M_{\mu}(\beta;\alpha,a[k,\lambda))  &  &   \label{areduc}
\eea 
where
\bdm
 K_{\mu \nu}= g_{\mu\nu}(\partial^2 + M^2)+(\frac{1}{\xi}-1)\partial_
    {\mu}\partial_{\nu}     \label{Kuv}
\edm
has the properties
\bdm
 K_{\mu \rho}d^{\rho \nu}(x)=\delta^{\nu}_{\mu}\delta^4(x), \;\;\;\;\;\;
 \partial^{\mu}K_{\mu \nu} = (\frac{1}{\xi}\partial^2 + M^2)\partial_{\nu} 
        \label{dK} 
\edm
For longitudinal polarization
\bdm
  \varepsilon^{\mu}(k,\lambda_{\ell})= (\frac{k}{M},\hat{k}\frac{\omega}{M}) 
        =\frac{k^{\mu}}{M} +\Delta^{\mu}, \;\;\;\;\Delta^{\mu}={\cal O}(
\frac{M}{\omega})   \label{longpol} 
\edm
So
\bdm
   \M_{\ell}(\cdots)=\varepsilon^{\mu}(k,\lambda_{\ell})\M_{\mu}(\cdots) \\
  =\frac{k^{\mu}}{M}M_{\mu}(\cdots)(1+{\cal O}\frac{M^2}{\omega^2})
        \label{approx}
\edm
Then recalling that $A^{\star} = A$ and using (\ref{dK}) and (\ref{0F})
\bea
  i(2\pi)^4\delta^4(\cdots)\frac{\km}{M}\M_{\mu}(\beta;\alpha,a(k,\lambda_{\ell})) 
   =\frac{(-i)^2}{M}\int\;(dx)e^{-ikx}(\frac{1}{\xi}\partial^2 + M^2)
   <\beta|\partial_{\nu}A^{\nu^{\star}}(x)|\alpha>  &  &  \nonumber  \\
      = \int\;(dx)e^{-ikx}(\partial^2 + \xi M^2)<\beta|\chis(x)|\alpha> 
       =i(2\pi)^4\delta^4(\cdots)(-i)\M(\beta;\alpha,\chi(k)) &  &  \label{ET1}
\eea
This provides the result
\bdm
  \frac{\km}{M}\M_{\mu}(\beta;\alpha,\akl)=-i\M(\beta;\alpha,\chi(k))
       \label{ET1a}
\edm
and the high energy limit for a longitudinaly polarized vector boson
\bdm
  \M(\beta;\alpha,\akl)=-i\M(\beta;\alpha,\chi(k))(1+{\cal O}\frac{M^2}{
     \omega^2})   \label{ET1b}
\edm
Note that the LSZ derivation specifies that in (\ref{ET1a}),(\ref{ET1b}) 
the k in $a(k,\lambda_{\ell})$ and the k in $\chi(k)$ are the same four-
vector. From (\ref{masses}) we see that for arbitrary $\xi$ this can 
only be satsfied in the limit $M^2=0$. The one important exception \cite{hv}  
is the Feynman gauge $(\xi=1)$ for which the vector boson and the wouldbe 
Goldstone boson are on the same mass shell, $k^2=M^2$. We also note that 
even in Feynman gauge, the left hand side of (\ref{ET1a}) is  an 
S-Matrix element only in the limit because it is not on the ``spin-1 shell''. 
A term proportional to $k_{\mu}$ in $\M_{\mu}$ will contribute a term 
proportional to $M^2$ to $k^{\mu}\M_{\mu}$, but zero to the S-Matrix element 
$\varepsilon^{\mu}\M_{\mu}$.

 For two external vector bosons, the LSZ manipulations are a little more 
involved, but again follow from (\ref{0F}),(\ref{dK}), without requiring 
any referennce to Ward identities involving ghost propagators 
or vertices. Start from
\bea
 \idelt\frac{k_2^{\nu}}{M}\frac{k_1^{\mu}}{M}\M_{\mu \nu}
  (\beta;\alpha,a(k_2,\lambda_{\ell}),a(k_1,\lambda_{\ell}))  
  \hspace{1.7in}                   &  &  \nonumber  \\ 
 =\frac{k_2^{\nu}}{M}\frac{k_1^{\mu}}{M}(-i)^2\int\int\;(dx')(dx)
 e^{-i(k_2 x'+k_1 x)}K'_{\nu \tau}K_{\mu \sigma}
 <\beta|T(A^{\tau}(x')A^{\sigma}(x))|\alpha> &  &  \label{start}
\eea
The k's are converted to derivatives on the A's by partial integration 
from the exponentials and use of (\ref{dK}). (We neglect possible 
contact terms which don't contribute to the S-matrix). Each $\partial A$
is replaced by $F-X$ where $X\equiv \xi M \chis$ (\ref{F4}). This generates 
four terms. 
The FF term is proportional to the matrix element
$$
  <\beta|T(F((x')F(x))|\alpha> = -\xi<\beta|T(\{Q,\bar{c}(x')\}F(x))|\alpha> 
$$ 
which vanishes
because the BRST charge Q commutes with F and anihilates the physical states 
 $<\beta|$ and $|\alpha>$.
An FX term is of the form
\bea
\frac{1}{M} \int\;(dx')e^{-ik_2 x'}(\frac{1}{\xi}\partial'^2 +M^2) 
      \hspace{1.7in}       \nonumber  \\
\times\int\;(dx)e^{-ik_1 x}(\partial^2 +\xi M^2)<\beta|T(F(x')\chis(x))|
\alpha>                                
    \nonumber
\eea
The integral on the second line is just the reduction formula for $\chis$.
Run back it is equal to 
$$  -i<\beta|F(x')|\alpha,\chi(k_1)> = 0 $$
The XX term is
\bea
  -i^2\int\;(dx')\int\;(dx)e^{-i(k_2 x' +k_1 x)}
 (\partial'^2 +\xi M^2)(\partial^2 +\xi M^2)
 <\beta|T(\chis(x')\chis(x))|\alpha> &  &   \nonumber  \\
  =-\idelt\M(\beta;\alpha,\chi(k_2),\chi(k_1)>  \label{XX}
\eea
Matching this to the start of (\ref{start}) gives the result 
\bdm 
 \frac{k_2^{\nu}}{M}\frac{k_1^{\mu}}{M}\M_{\mu \nu}(\beta;\alpha,
   a(k_2,\lamlong),a(k_1,\lamlong))= -\M(\beta;\alpha,\chi(k_2)\chi(k_1))
        \label{M2}
\edm
and the high energy limit for two longitudinaly polarized vector bosons
\bdm
  \M(\beta;\alpha,a(k_2,\lambda_{\ell}),a(k_1,\lambda_{\ell})) = 
    -\M(\beta;\alpha,\chi(k_2),\chi(k_1))(1+{\cal O}\frac{M^2}{\omega^2})
        \label{M2a}
\edm
The remarks made after (\ref{ET1a}),(\ref{ET1b}) apply as well to 
(\ref{M2}),(\ref{M2a}).
  At first sight, this appears to say that the \ET holds 
without modification  order by order 
in the perturbative loop expansion. But we still have to relate the 
renormalized $\tsf s \; <0|T(\phis \cdots)|0>$ appearing in the LSZ reduction 
formulas to the renormalized $\tf s \;  <0|T(\phi \cdots|0>$ calculated 
according to section two, to maintain the BRST structure defined in the 
unrenormalized theory. This is ($\phi$ is any field in the theory)
\bdm
  \tau^{\star}(\cdots)=<0|T(\phis \cdots)|0> =\sqrt{\frac{Z_{\phi}}{
Z^{\star}_{\phi}}}\cdots <0|T(\phi \cdots)|0>  \label{ts t}  
\edm 
So when the perturbative calculation of the S-matrix elements (the ${\cal M}$'s)
is given in terms of $\tf$s, a factor $\sqrt{\frac{Z_{\phi}}{Z^{\star}_{\phi}}}
$ will appear for each $\phi$ which is not star-normalized, ($Z_{\phi}\neq 
Z^{\star}_{\phi}$). In particular, in the renormalization scheme of section 
two, a single Z is used to renormalize both $h_0$ and $\chi_0$. In the 
case of \ssb, $Z^{\star}_h \neq Z^{\star}_{\chi}$ so it is not possible for 
both $\hat{h}$ and $\chi$ to be star-normalized. There is yet another 
factor which has to be taken into acount. For S-matrix elements computed 
from $\tsf$s, there are no loop corrections to external lines. LSZ amputation 
by the inverse free propagator removes the complete on-shell external 
propagator because it has unit residue at its pole. When S-matrix elements are 
computed from $\tf$s, there are loop corrections to external lines which 
survive the LSZ amputation. These are of the form
\bea
  -i(p^2-m^2)\{ \frac{i}{p^2-m^2} +\frac{i}{p^2-m^2}(-i\Sigma^{FD}(p^2)+ 
 i\delta Z(p^2-m^2)-\delta m^2)\frac{i}{p^2-m^2} +\cdots\}\underline{
\tilde{\tau}}  &   &  \nonumber  \\
 =\{1 +(\frac{\partial\Sigma^{FD}}{\partial p^2}-
\delta Z)\}\underline{\tilde{\tau}}
 =\{1+\delta Z^{\star}
 -\delta Z + \cdots \}\underline{\tilde{\tau}} = \frac{Z^{\star}}{Z} 
 \underline{\tilde{\tau}} \hspace{1in} &   &    \label{amp}
\eea
This factor just reverses that of (\ref{ts t}) So the final result is
\bdm
  {\cal M}(\hat{h}\cdots A_t \cdots A_{\ell}\cdots)= (-i)^{n_{\ell}}
 (\frac{Z^{\star}_h}{Z})^{\frac{n_h}{2}}(\frac{Z^{\star}_{\chi}}{Z})
 ^{\frac{n_{\ell}}{2}}\underline{FT<0|T(\hat{h}\cdots \varepsilon\cdot A_t 
\cdots \chi \cdots)|0>}
 \{1 + {\cal O}(\frac{M^2}{\omega^2})\}               \label{result}
\edm
The FT is fourier transform with four-momentum delta function factored off; 
and the underline means complete external line propagators removed.
\bdm
     \tilde{\tau}(\cdots)_n = (iD)^n\underline{ \tilde{\tau}}(\cdots)_n 
       \label{AMP}
\edm
The result (\ref{result}) is simple, intuitive, and quite general.    
(If the vector boson two-point function is not normalized to unit 
residue ,(\ref{result}) also includes $\sqrt{\frac{Z_A^{\star}}{Z_A}}$ factors).
For $k^{\mu}k_{\mu}= M^2$, the left hand side is an 
S-matrix element and must be independent of $\xi$. 
In the high energy limit, 
in the leading order, there is no contribution from the  
${\cal O}( M^2/\omega^2)$ terms on right hand side  and in this leading order, 
 the product of ratios of Z factors and the 
 amputated $\tf$ (which product is just the on-shell $\underline{\tau}^{\star}$ 
function) 
must be independent
of $\xi$. In section five we carry out an explicit one-loop perturbative 
calculation to check that the correction factors in (\ref{result}) are 
indeed correct.

There is still some freedom of choice of \rcs to complete the specification 
of the renormalization (\ref{R}).  
 To conclude this section, we recall the choices we have made  
in order to arrive at the results (\ref{M2}),(\ref{result}). (1)The gauge 
fixing function F is chosen to cancel tree level $\chi-\partial A$ mixing.
($\kappa= 1+\mbox{one-loop}$). (2)Renormalization (\ref{R}) is constrained 
to maintain BRST symmetry in same form as unrenormalized. (3)Renormalized 
masses are the physical masses (modulo unstable particles. See discussion 
below). (4)The vector boson A is renormalized to satisfy the unit residue 
condition ($Z_A =Z^{\star}_A$). (5)A specific choice of the second gauge 
parameter $\kappa$ is made to get (\ref{M2}) to come out with no extra 
factors. Since the end results (\ref{M2a}),(\ref{result}) are S-matrix 
elements, presumably other renormalization schemes are admissable, but the 
one outlined here seems the simplest and most physical and intuitive. 
(See also the example 
explicit calculation in section five, and also of course, alternative 
treatments from \cite{hky} and references cited therein).

\section{\bf Decay of an Unstable Particle and the Equivalence Theorem}

 Let a,$\chi$ be generic vector bosons and wouldbe \GB, with masses squared
$M^2$ and $\xi M^2$ and $m=m_h$ the Higgs boson mass.
  If $m_h > 2 m_a$, the Higgs boson is unstable and the standard calculation 
of the decay rate for $h\rightarrow a_{\ell} \;a_{\ell}$ is given by 
\bdm
  \Gamma = \frac{1}{2m}|{\cal M}(h;a(k_1,\lambda_{\ell}),a(k_2,\lambda_{\ell})) 
                      |^2 \frac{1}{2}\int\;d \Phi  \label{gamma}
\edm
If the \ET applies, (In this case, the high energy limit is just the mass of 
Higgs boson much greater than the 
masses of the gauge bosons or wouldbe \GB)
 one may substitute for ${\cal M}$ in (\ref{gamma}) the invariant matrix-
element for decay into two wouldbe \GB.

 Since the h particle passes from stable
 to unstable by an arbitrarily small change in 
the continuous parameter $m^2/M^2 = 2\lambda/g^2$ from $< 4$ to $>4$,
continuity suggests that the form of $\M$  should be unchanged 
as the pole of the h propagator moves continuously into the second sheet.
\cite{ELOP}

A derivation based on unitarity of the resonance contribution to elastic 
 scattering goes as follows.  
With $M^2,\xi M^2 < m^2$,the vector boson and the wouldbe 
Goldstone boson are stable. So we can apply the \ET 
to the elastic scattering processes $a_{\ell}+a_{\ell}\ra a_{\ell}+a_{\ell},
\;\;a_{\ell}+\chi \ra a_{\ell}+\chi,\;\;\chi+\chi \ra \chi+\chi$ in the limit
$M^2 \ll \omega_i^2$.

Unitarity for the invariant matrix element $\M $ for forward elastic 
scattering of two identical spinless particles, $\chi+\chi\ra \chi+\chi $,is
\bea
 2\;\Im\M_{\alpha \alpha}\;=\;\frac{1}{2}\int\;d\Phi_2|\M_{\alpha\alpha}|^2 
&   & (\alpha = \chi\;\chi)  \nonumber    \\
  &  d\Phi_2 = \frac{1}{16\pi^2}\frac{p}{W}d\Omega  \label{unit}
\eea
For a narrow resonance
\bea
 \M = \M_{res}+ \M_{bgd}  &   &  \nonumber  \\
 \M_{res}= - \frac{R^2}{s - m_p^2 + i m_p\Gamma}  \label{res}
\eea

Consider summed renormalized perturbation theory.(skeleton expansion)
~
~
\begin{center}\begin{picture}(600,300)(0,0)

\CArc(60,160)(40,0,360)
\dl(15,215)(30,190){3}
\dl(15,105)(30,130){3}
\dl(105,215)(90,190){3}
\dl(105,105)(90,130){3}
\SetOffset(30,0)
\CArc(200,160)(15,0,360)
\Line(200,175)(200,190)
\Line(200,145)(200,130)
\Boxc(200,202)(24,24)
\Boxc(200,118)(24,24)
\dl(188,214)(178,224){3}
\CArc(170,232)(10,0,360)
\dl(162,240)(152,250){3}
\dl(212,214)(222,224){3}
\CArc(230,232)(10,0,360)
\dl(238,240)(248,250){3}
\dl(188,106)(178,96){3}
\CArc(170,88)(10,0,360)
\dl(162,80)(152,70){3}
\dl(212,106)(222,96){3}
\CArc(230,88)(10,0,360)
\dl(238,80)(248,70){3}
\Text(130,160)[]{$=$}
\Text(290,160)[]{$ +\cdots$}
\end{picture}  \\  
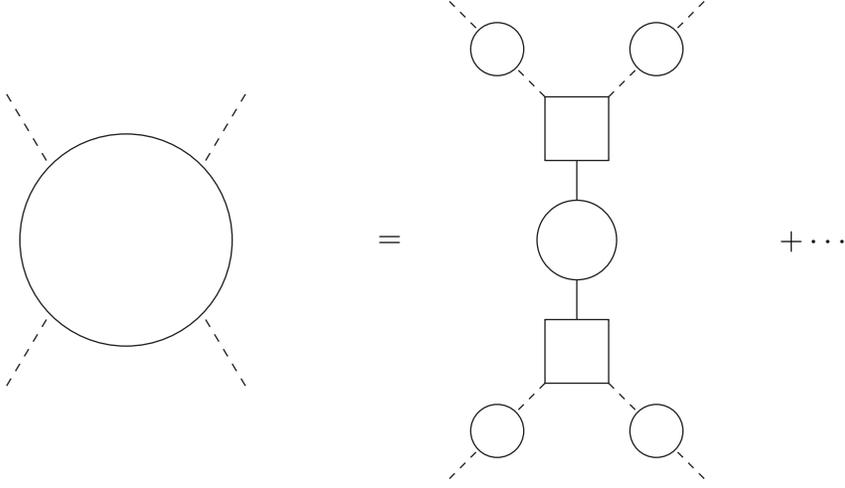
{Fig.1. Resonance at complex pole of summed propagator of unstable particle} 

\end{center}
~

\bdm
\tau= (iD_{\chi})^4 (iV)^2 iD_h = (\frac{Z^{\star}_{\chi}}{Z_{\chi}})^2
 \tau^{\star} =
(\frac{Z^{\star}_{\chi}}{Z_{\chi}})^2 (iD^{\star}_{\chi})^4(iV^{\star})^2
 iD^{\star}_h
       \label{A3}
\edm
V is the completely amputated 1PI $h\;\chi\;\chi$ vertex function (diagrams 
in the third columns of figures three and four). 
\bdm
 V = \frac{Z_{\chi}}{Z^{\star}_{\chi}}
(\frac{Z_h}{Z^{\star}_h})^{\frac{1}{2}} V^{\star}
     \label{V}
\edm
The summed h-propagator is  
\bdm
 D_h = \frac{Z^{\star}_h / Z_h}{k^2-m_r^2- \Re\bar{\bar{\Sigma_h}}^{\star}(k^2)
     -i\Im \Sigma_h^{\star} (k^2)}  \label{Dh}
\edm
The double bar indicates two subtractions (in addition to multiplicative 
renormalization) for $\Re\Sigma$. The first subtraction secures $\Re\Sigma 
(m_r^2)=0$.  The second subtraction makes the derivative of 
$\Re\Sigma(k^2)$ vanish at $k^2 = m_r^2$.  

We observe that the renormalized mass $m_r^2$ defined by these subtractions 
is not the same as the mass $m_p^2$ which is a parameter of the exact pole 
position in (\ref{res}). The difference is
$$
    m_p^2 - m_r^2 ={\cal R}e(\Re \bar{\bar{\Sigma}}(s_p)+i\Im \Sigma(s_p)) 
$$
This difference is second order in the expansion parameter of the 
theory
$$
     g^2\frac{m^2}{M^2},\;\;\;\;\;\;(\frac{\Gamma}{m}\sim g^2\frac{m^2}{M^2})
$$
So the difference between these two renormalized masses can be ignored in
a first order perturbative calculation. But the difference between the 
complex pole position and the real mass squared is a first order imaginary 
quantity ($-i m \Gamma$) which will be essential for the verification of the 
\ET .

Applying LSZ, the resonance contribution to the invariant matrix element 
for $\chi+\chi \rightarrow \chi+\chi$ is
\bea
 i\M_{res}  = &   iV^{\star} iD^{\star}_h iV^{\star} = &
          (\frac{Z^{\star}_{\chi}}{Z_{\chi}})^2 iV iD_h iV  \nonumber  \\
&= i\frac{-(\frac{Z^{\star}_{\chi}}{Z_{\chi}}\sqrt{\frac{Z^{\star}_h}{Z_h}}V)^2}
{s-m_r^2-i\Im \Sigma_h^{\star} (m_r^2)} &  \equiv i\frac{-(f V)^2}{s -m_r^2 + i
m_r \Gamma}     \label{mres}
\eea 

Substitute this into the unitarity equation (\ref{unit})
\bea
  2\;\Im \M_{\alpha \alpha} =2 \frac{(f V)^2m_r\Gamma}{(s-m_r^2)^2 +
m_r^2 \Gamma^2}  &  &  \nonumber \\
 = \frac{1}{2}\int\;d \Phi_2|\M_{\alpha \alpha}|^2 =\frac{1}{2}\int\;
d\Phi_2\frac{(fV)^4}{(s-m_r^2)^2 + m_r^2 \Gamma^2} & &  \label{unit2}
\eea
which implies
\bdm
  \Gamma = \frac{1}{2 m_h}|f V|^2 \frac{1}{2}\int\;d \Phi_2
       \label{proof}
\edm
Repeating the unitarity calculation for the other two initial/final
states,$\chi\;a_{\ell},\;\;a_{\ell}\;a_{\ell}$  gives two more versions of 
(\ref{proof}),each with f being the  product of ratios of \rczs required 
to convert amputated $\tf$s to \SMEs.  
 Through \1l the perturbabtive calculation of the 1PI  vertex function  
encounters no multiple poles (no Dyson resummation required). 
 And for the calculation 
of the Higgs decay rate, the real part of the variable $ k^2$ is 
 fixed at $m_r^2$ 
i.e. at the resonance peak; so the calculation does not depend on choice 
of off-peak representation (e.g. partial wave threshold factors) of the
 resonant scattering amplitude.

  To check the applicability of the \ET to the calculation of the 
decay rate of an unstable particle 
, the tree plus one-loop 
perturbative calculations of the 1PI vertex functions $V_{h,a_{\ell},
a_{\ell}},\;V_{h,a_{\ell},\chi},\;V_{h,\chi,\chi}$ are given in the next 
sections.

\section{\bf Perturbative Calculation in the Abelian Higgs model}

We complete the specification of the renormalization 
by enumerating the parameters and \rcs and \rczs.

The renormalized parameters (\ref{R}) on which physical quantities can 
depend are 
\bdm
       \mu^2,\;\;\lambda,\;\;g    \label{params}
\edm
There are two gauge parameters
\bdm
      \xi, \;\; \kappa     \label{gparams}
\edm
The gauge parameter $\kappa$ has been fixed by considerations given above 
, equation (\ref{K}).The gauge parameter $\xi$ is left free.
The corresponding \rczs are
\bdm
  Z_{\mu^2},\;\;Z_{\lambda},\;\;Z_g,\;\;Z_{\xi},\;\;Z_{\kappa} \label{rcz1}
\edm
The field strength \rczs are
\bdm
    Z_A = Z^{\star}_A,\;\;Z   \label{fsrcz}
\edm

$Z_{\kappa}$ is fixed for consistent renormalization of the gauge fixing 
    function F
  (\ref{ZK}),  $\;Z_{\xi},Z_g$ are related to $Z_A$ by Ward identites (\ref{WI}).
There is also V, which has to be determined as a function of the 
renormalized parameters of the theory (including $\xi$). We introduce 
$\zeta_v$ by the relation
\bdm
    V = \zeta_v v  = v + (\zeta_v -1)v   \label{zeta}
\edm 
\bdm
           v^2 = \frac{(-\mu^2)}{\lambda},\;\;\;\; 
              \zeta_v=1+\zeta_v^{\mbox{one-loop}}+\cdots  \label{zeta2}
\edm
We trade $\mu^2\;,\lambda,\;g$ for $m,\;M,\;g$
\bdm
  m^2 = 2(-\mu^2),\;\; M^2 = g^2\frac{(-\mu^2)}{\lambda},\;\;
           (\lambda = \frac{g^2}{2}\frac{m^2}{M^2})  \label{params2}
\edm

  So we require five conditions to fix the four \rczs $Z_{\mu^2},\;
   Z_{\lambda},\;Z_A,\;Z \;$ and $\zeta_v$ (the vev).

(1)All tadpoles are canceled order by order
$$
       <\hat{h}> = iT = 0
$$
(2)The induced $\chi-\partial A$ mixing vanishes at $k^2=\xi M^2 =M^2_{\chi}$
$$
      X^{\mu} = k^{\mu}X(k^2), \;\;\;\;\;\; X(\xi M^2)= 0
$$
(3),(4)The gauge invariant part of the complete renormalized vector boson 
propagator has pole at the physical mass with unit residue.
$$
    \Pi_T(M^2)= 0, \;\;\;\; \frac{d \Pi_T(M^2)}{d k^2}= 0
$$
(5) The stable h propagtor has pole at the physical mass.
$$
 \Sigma_h(m^2) = 0, \;\;\;\;\;\mbox{for unstable h}\;\;\;\;\Re\Sigma_h(m_r^2)=0
$$
Note that neither condition,$\frac{d\Sigma_h(m^2)}{d k^2}= 0$ or 
$\frac{d\Sigma_{\chi}(\xi M^2)}{d k^2}= 0$ is included. Thus Z is not 
required to be equal to either $Z_h^{\star}$ or $Z_{\chi}^{\star}$.

In these equations, each of the functions $T,\;X,\;\Pi_T,\;\Sigma_h\;$ is a
renormalized $1PI$ function. Each has two components
$$
                 f = f^{FD} + f^{ct}
$$
The first is the sum of (one-loop) Feynman diagrams with renormalized masses 
in the free propagators and renormalized coupling constants at the elementary 
vertices. The Feynman  integrals are evaluated with dimensional regularization. 
The second is the corresponding tree diagram multiplied by $(Z_i -1)$ and 
$(\zeta_v -1)$ factors.

The \rczs and $\zeta_v$ which are fixed by these conditions are given in 
appendix A.  

(5')With no adjustable parameter
$$
  \Sigma_{\chi}(\xi M^2)_{M^2\rightarrow 0} 
\rightarrow 0 
$$
The limit $M^2\rightarrow 0,\;g^2\rightarrow 0,\;v^2=\frac{-\mu^2}
{\lambda}$ fixed is the "gaugeless" limit in which the broken symmetry 
is a global symmetry and $\chi$ is a genuine Goldstone boson.  Since we 
have been careful to renormalize consistent  with the unbroken symmetry,
the Goldstone boson remains massless after renormalization.\cite{bw}
We see that in the context of the Abelian Higgs model, the gaugeless 
limit could as well be called the Goldstone limit. Furthermore, the 
\ET is supposed to hold in the limit of energies much greater than 
the gauge boson masses. For the decay of the Higgs, the energy is the
Higgs mass and the condition $\frac{M^2}{m^2}\ll 1$ is just this limit.
The result $\delta\zeta_v = 0$ is special to this limit, in which the
Goldstone theorem provides for the same counter term to eliminate 
both the tadpoles and the Goldstone boson mass. When the gauge 
interactions are included, $\delta\zeta_v$ is both gauge dependent 
and ultraviolet divergent.

   We turn now to the explicit calculation of the decay probabilites 
(\ref{gamma}) for $ h\rightarrow a,\;a,\;\;h\rightarrow a,\;\chi,\;\;
h\ra \chi,\;\chi$. For these processes, $\omega$ is $m/2$. To check the
\ET,we compute to leading order in powers of $M^2/m^2$. In this order,
the longitudinal polarization vectors are replaced by $k^{\mu}$/M .
At \tl, the invariant matrix elements (the $ \M' s$) are equal to the 
\tl $\underline{\tilde{\tau}}'s$ which are just the elementary vertices 
contracted with $k^{\mu}$/M for each external vector boson.(Fig.2.) (The ratios 
of Z factors differ from one only at one-loop order).

\begin{center}\begin{picture}(600,70)(0,0)

\Line(20,40)(70,40)
\zz(70,40)(90,60){2}{3}  \Line(90,50)(75,60)
\zz(70,40)(90,20){2}{3}  \Line(90,30)(75,20)
\Line(190,40)(240,40)
\dl(240,40)(260,60){3}
\zz(240,40)(260,20){2}{3} \Line(260,30)(245,20)
\Line(360,40)(410,40)
\dl(410,40)(430,60){3}
\dl(410,40)(430,20){3}

\Text(110,40)[]{\small 2.0}
\Text(280,40)[]{\small 1.0}
\Text(450,40)[]{\small 0.0}
\Text(10,40)[]{-i}
\Text(350,40)[]{i}

\end{picture} \\
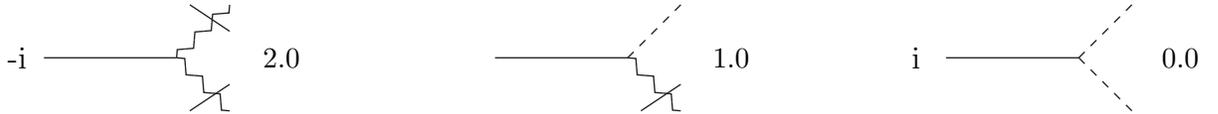
{ Fig.2.Tree diagrams for $h\rightarrow a_{\ell}a_{\ell},
\chi a_{\ell},\chi \chi$.  \linebreak
 The slashes indicate contraction with $k^{\mu}/M$.} 

\end{center}

  For decay into two longitudinal vector bosons:

\bdm
-i \frac{k_2^{\nu}}{M}\frac{k_1^{\mu}}{M}\underline{\tilde{\tau}}_{\mu \nu}
= 2 g M \frac{k_1\cdot k_2}{M^2}  \label{tree1}
\edm

But in the limit $M^2\rightarrow 0$,
$$ 
  2 k_1\cdot k_2 = K^2,
$$
the square of the four-momentum of Higgs boson, which is to be evaluated 
at its complex on-shell value,
$$
  K^2 = m^2 -i m \Gamma
$$
Substituting the zeroth order $ m^2 $ into (\ref{gamma})gives the first 
approximation to $\Gamma$
\bdm
     \Gamma = \frac{g^2}{32 \pi}\frac{m^3}{M^2}   \label{gamma2}
\edm
Substitute all this into (\ref{tree1})to obtain
\bdm
  -i \frac{k_2^{\nu}}{M}\frac{k_1^{\mu}}{M}\underline{\tilde{\tau}}_{\mu \nu}
 = g\frac{m^2}{M}-i\pi\frac{g^3}{16 \pi^2}\frac{m^4}{M^3}\frac{1}{2}.
                  \label{tree2}
\edm
For decay into one vector boson and one wouldbe Goldstone Boson
\bdm                        
 \frac{k_2^{\nu}}{M}\underline{\tilde{\tau}}_{\nu} = 
 \frac{k_2^{\mu}}{M}g (2 k_1-k_2)_{\mu}= 2 g \frac{k_1^{\mu}\cdot k_2^{\mu}}
   {M}   \label{tree3}
\edm
which is identical to (\ref{tree1}). For decay into two wouldbe \GB, the 
tree-level invariant matrix-element is simply
\bdm
   i\underline{\tilde{\tau}}= g\frac{m^2}{M}.   \label{tree4}.
\edm
There is no explicit dependence on the Higgs boson four-momentum; hence 
no imaginary shift is generated.

  For two transversely polarized vector bosons the tree amplitude is just
$g M$, down by $M^2/m^2$. Note that in the "gaugeless" limit, $g, M 
\ra 0$ while $m^2\frac{g}{M}$ remains finite. At one-loop order, the 
dominant terms are $g^3 m^4/M^3$, finite in the limit. Subdominant terms 
are of order $g^3 m^2/M$ and vanish in the limit. In this limit - which 
is equvialently the heavy Higgs limit - the trilinear and quartic scalar 
couplings are dominent over the gauge couplings. All one-loop diagrams with
internal gauge vector boson lines are in the subdominent class. This leads 
to substantial simplification of the calculation, and, along with the  
observation that the $\chi$ mass squared $\xi M^2$ also disapears in the limit, 
shows that there is no source of $\xi$ dependence in the calculation in 
this limit.

In figures 3 and 4 are shown the one-loop Feynman diagrams and counterterm
diagrams which contribute in this 
limit to the $\underline{\tilde{\tau}}$ functions whose tree-level 
values are given in (\ref{tree2},,\ref{tree4}). The corresponding 
 calculated values 
are given
in Table 1. 
~

\begin{center}\begin{picture}(600,410)(0,0)

\SetOffset(0,400)

\Line(10,40)(30,40)
\CArc(50,40)(20,0,360)
\zz(70,40)(90,60){2}{3}  \Line(75,60)(85,45)
\zz(70,40)(90,20){2}{3}  \Line(75,20)(85,35)
\Text(120,40)[]{\small 2.a}

\Line(360,40)(380,40)
\CArc(400,40)(20,0,360)
\dl(420,40)(440,60){3}
\dl(420,40)(440,20){3}
\Text(470,40)[]{\small 0.a}

\SetOffset(0,300)

\Line(10,40)(30,40)
\dcarc(50,40)(20,0,360){3}
\zz(70,40)(90,60){2}{3}  \Line(75,60)(85,45)
\zz(70,40)(90,20){2}{3}  \Line(75,20)(85,35)
\Text(120,40)[]{\small 2.b}

\Line(360,40)(380,40)
\dcarc(400,40)(20,0,360){3}
\dl(420,40)(440,60){3}
\dl(420,40)(440,20){3}
\Text(470,40)[]{\small 0.b}

\SetOffset(0,190)

\Line(360,40)(380,40)
\CArc(395,55)(20,40,220)
\dcarc(395,55)(20,220,40){3}
\dl(412,65)(430,80){3}
\dl(380,40)(410,15){3}
\Text(460,40)[]{\small 0.c}
\Text(340,45)[]{[2]}

\SetOffset(0,90)

\Line(10,40)(30,40)
\Line(30,40)(60,70)
\Line(30,40)(60,10)
\dl(60,10)(60,70){3}
\zz(60,70)(70,80){2}{3}  \Line(60,80)(70,70)
\zz(60,10)(70,0){2}{3}   \Line(60,0)(70,10)
\Text(100,40)[]{\small 2.d}

\Line(190,40)(210,40)
\Line(210,40)(240,70)
\Line(210,40)(240,10)
\dl(240,10)(240,70){3}
\dl(240,70)(250,80){3}
\zz(240,10)(250,0){2}{3}  \Line(240,0)(250,10)
\Text(270,40)[]{\small 1.d}

\Line(370,40)(390,40)
\Line(390,40)(420,70)
\Line(390,40)(420,10)
\dl(420,10)(420,70){3}
\dl(420,70)(430,80){3}
\dl(420,10)(430,0){3}
\Text(450,40)[]{\small 0.d}

\SetOffset(0,-10)

\Line(10,40)(30,40)
\dl(30,40)(60,70){3}
\dl(30,40)(60,10){3}
\Line(60,10)(60,70)
\zz(60,70)(70,80){2}{3}  \Line(60,80)(70,70)
\zz(60,10)(70,0){2}{3}   \Line(60,0)(70,10)
\Text(100,40)[]{\small 2.e}

\Line(190,40)(210,40)
\dl(210,40)(240,70){3}
\dl(210,40)(240,10){3}
\Line(240,10)(240,70)
\dl(240,70)(250,80){3}
\zz(240,10)(250,0){2}{3}  \Line(240,0)(250,10)
\Text(270,40)[]{\small 1.e}

\Line(370,40)(390,40)
\dl(390,40)(420,70){3}
\dl(390,40)(420,10){3}
\Line(420,10)(420,70)
\dl(420,70)(430,80){3}
\dl(420,10)(430,0){3}
\Text(450,40)[]{\small 0.e}

\end{picture}   \\   \vspace{.2in}

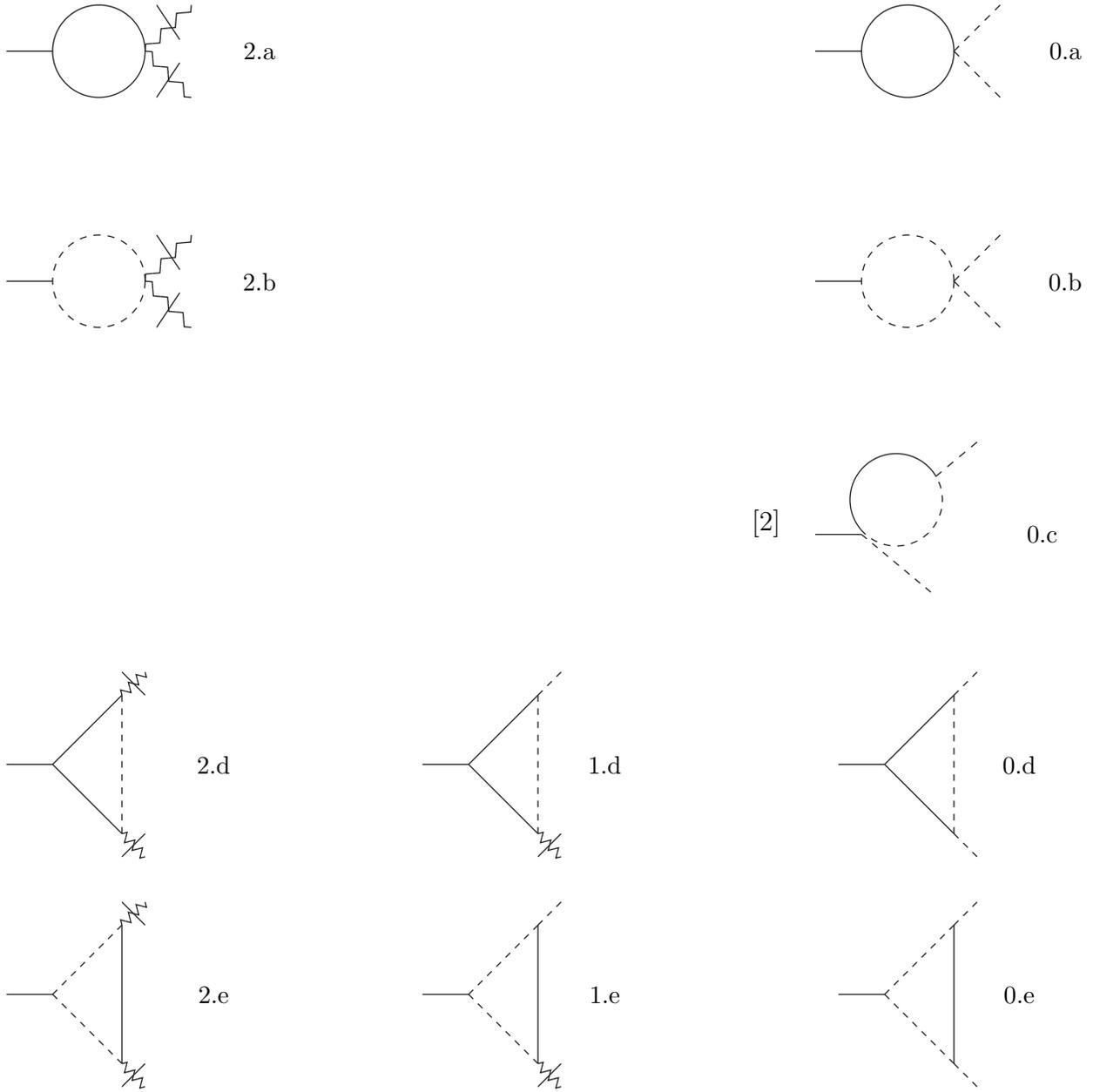
{Fig.3. One-loop diagrams which contribute to $h \rightarrow a_{\ell}
   a_{\ell},\chi a_{\ell},\chi \chi$ in the limit $g\rightarrow 0,M
  \rightarrow 0,g/M = 1/v$. In this limit there are no contributions 
  2.c,1.a,1.b,1.c}

\end{center}
\begin{center}\begin{picture}(600,60)(0,0)

\Line(20,40)(70,40)
\zz(70,40)(90,60){2}{3}  \Line(90,50)(75,60)
\zz(70,40)(90,20){2}{3}  \Line(90,30)(75,20)
\CArc(70,40)(4,0,360)
\Line(190,40)(240,40)
\dl(240,40)(260,60){3}
\zz(240,40)(260,20){2}{3} \Line(260,30)(245,20)
\CArc(240,40)(4,0,360)
\Line(360,40)(410,40)
\dl(410,40)(430,60){3}
\dl(410,40)(430,20){3}
\CArc(410,40)(4,0,360)

\Text(10,40)[]{-i}
\Text(350,40)[]{i}
\Text(40,0)[]{$(Z_{4g}\zeta_v -1)$}
\Text(210,0)[]{$(Z_{3g} -1)$}
\Text(385,0)[]{$(Z_{4\lambda}\zeta_v -1)$}

\end{picture}  \\   \vspace{.2in}

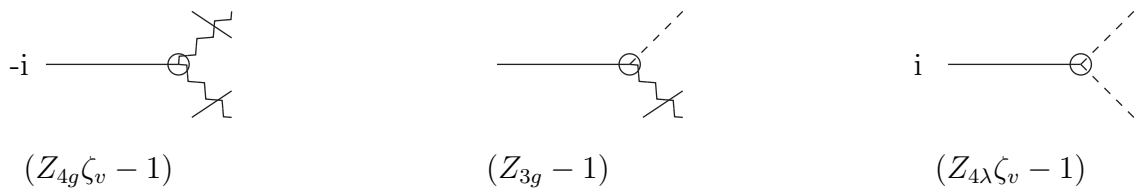
{Fig.4. One-loop counterterm diagrams. Each factor displayed is multiplied 
by common factor $g\frac{m^2}{M}$}  \hspace{3in}

\end{center}

\begin{table}
\begin{center}

\begin{tabular}{|c|c|c|c|}
   &
$-i\frac{k_2^{\nu}}{M}\frac{k_1^{\mu}}{M}\underline{\tilde{\tau}}_{\mu \nu}$ &
$\frac{k_2^{\nu}}{M}\underline{\tilde{\tau}}_{\nu}$ &
$ i\underline{\tilde{\tau}}$  \\  [.1in]\hline 
 a   &$ -\frac{3}{2}\Delta_{\epsilon} -3 +\frac{\sqrt{3}}{2}\pi$ &  & 
$ -\frac{3}{2}\Delta_{\epsilon} -3 +\frac{\sqrt{3}}{2}\pi$ \\
 b  &$ -\frac{1}{2}\Delta_{\epsilon} -1 +i\pi(-\frac{1}{2})$  &  & 
$ -\frac{3}{2}\Delta_{\epsilon} -3 +i\pi(-\frac{3}{2})$  \\
 c  &   &  &$ 2[-\De -1]$  \\
 d  & $\frac{3}{2}\De +6 -\frac{3\sqrt{3}}{2}\pi+\frac{\pi^2}{3}$ &
 $ 3 -\sqrt{3}\pi +\frac{1}{3}\pi^2$  & $\frac{\pi^2}{3}$ \\
 e  &$ \frac{1}{2}\De +\frac{\pi^2}{12}+i\pi(-\frac{1}{2}+\ln{2})$  &
  $-1+\frac{\pi^2}{12}+i\pi(-1+\ln{2})$ & 
  $\frac{\pi^2}{12}+i\pi(\ln{2})$  \\ 
 ct  & $-\frac{1}{2}$ & $-\frac{1}{2}$ & $5\De +\frac{19}{2}-\frac{3\sqrt{3}}{2}
    \pi$ 
\end{tabular}
\caption{The one-loop vertex contributions in the Abelian Higgs model.  
Each entry is multiplied by
  common factor  $\frac{g^3}{16\pi^2}\frac{m^4}{M^3}$} 
\end{center}
\end{table}

~

~

The result of adding all the contributions to each of these is
$$
-i\frac{k_2^{\nu}}{M}\frac{k_1^{\mu}}{M}\underline{\tilde{\tau}}_{\mu \nu}
= g\frac{m^2}{M}\{1+\g2\frac{m^2}{M^2}[\frac{3}{2}-\sqrt{3}\pi+\frac{5}
{12}\pi^2 +i\pi(\ln{2}-1)]+\cdots\}
$$
$$
\frac{k_2^{\nu}}{M}\underline{\tilde{\tau}}_{\nu} =
= g\frac{m^2}{M}\{1+\g2\frac{m^2}{M^2}[\frac{3}{2}-\sqrt{3}\pi+\frac{5}{12}
\pi^2 +i\pi(\ln{2}-1) +\cdots]\}
$$
$$
 i\underline{\tilde{\tau}}
= g\frac{m^2}{M}\{1+\g2\frac{m^2}{M^2}[\frac{3}{2}-\sqrt{3}\pi+\frac{5}{12}
\pi^2 +i\pi(\ln{2} -\frac{3}{2}) +\cdots]\}
$$
which differ in the imaginary parts. To get the $\M$'s to put into          (\ref{gamma}), we have 
to multiply by the factors of ratios of square roots of Z's required to
convert amputated $\tf$s, renormalized as described above, to \SMEs. 
 Using the ratios given in appendix A, these factors are
$$
  \sqrt{\frac{Z^{\star}_h}{Z}}= 1+\g2 \frac{m^2}{M^2}[\frac{11}{4} -
  \frac{\sqrt{3}}{2} \pi]
$$
$$
  \sqrt{\frac{Z^{\star}_h}{Z}}
\sqrt{\frac{Z^{\star}_{\chi}}{Z}}= 1+\g2 \frac{m^2}{M^2}[\frac{11}{4} -
  \frac{\sqrt{3}}{2} \pi]
$$
$$
   \sqrt{\frac{Z^{\star}_h}{Z}}
\frac{Z^{\star}_{\chi}}{Z}= 1+\g2 \frac{m^2}{M^2}[\frac{11}{4} -
  \frac{\sqrt{3}}{2} \pi]
$$
which are all the same because of the special circumstance that 
$Z=Z^{\star}_{\chi}$
And we have to include the imaginary parts generated from the tree-level
amplitudes by evaluating at the complex four-momentum squared (\ref{tree2})
When all the factors are combined, the three  
invariant matrix elements are now the same.
\bea
 \frac{k_2^{\nu}}{M}\frac{k_1^{\mu}}{M}\M_{\mu \nu}=  i\frac{k_2^{\nu}}
    {M}\M_{\nu} =   (-1) \M  \hspace{1.2in}  &  &  \nonumber  \\
 = g\frac{m^2}{M}\{1+\g2 \frac{m^2}{M^2}[\frac{17}{4}-\frac{3\sqrt{3}}{2}\pi
   +\frac{5}{12}\pi^2 +i\pi(\ln{2}-\frac{3}{2}) +\cdots]\}
            (1+{\cal O}\frac{M^2}{m^2}) &  &  \label{rah}
\eea

\section{\bf The calculation in the Standard Model}

   The BRST structure of the Nonabelian gauge sector of the \sm is somewhat 
more complicated than that of the Abelian Higgs model presented in section 
two. But the key result (\ref{0F}), used in sections three and four, is
maintained; so we go directly to the perturbative calculation

    We calculate the decay rate for $h \ra W_+\;W_-$ and $h\ra \phi_+ \;
\phi_-$ through one-loop order in the \sm, in the heavy Higgs limit,
$m^2\gg M^2 \;(\lambda \gg g^2)$. We also take the Weinberg angle to be zero. 
The leading terms in this limit can 
be calculated starting from a truncated \sm Lagrangian.
\bea
 {\cal L} =  \frac{1}{2}(\partial \hat{h})^2 + (\partial \phi^{\dag})
\partial \phi + \frac{1}{2}(\partial\phi_0)^2 -\frac{1}{2}m^2 \hat{h}^2
  \hspace{2in} &    &    \nonumber  \\
    -\lambda v\hat{h}(\hat{h}^2 +2 \phi^{\dag} \phi +\phi_0^2)
    -\frac{\lambda}{4}(\hat{h}^2 +2\phi^{\dag} \phi +\phi_0^2)^2 \hspace{2in}   
  &  &   \nonumber  \\
 +g M\hat{h}W^{\dag}W
  -i\frac{g}{2}(\phi^{\dag}\stackrel{\lra}{\partial}\hat{h})W
 -(\phi\oa {\partial}\hat{h})W^{\dag})-\frac{g}{2}((\phi\oa{\partial}\phi_0)
 W^{\dag} +(\phi^{\dag}\oa{\partial}\phi_0)W)  &  & \nonumber  \\
 + {\cal L}_{ct}  \hspace{4.5in}  &  & \label{trunc}
\eea

\bdm
  m^2 = 2 \lambda v^2, \hspace{.5in} M^2 = \frac{1}{4}g^2 v^2, \hspace{.5in} 
   \lambda = \frac{g^2}{8}\frac{m^2}{M^2}   \label{sm}
\edm

For the calculation of $h\ra \phi_+\phi_-$, the purely scalar Lagrangian 
is sufficient. The calculation of $h\ra W_+W_-$  requires also the 
couplings to the ``external''W field with fixed coupling $g$.

  The tree-level for decay into longitudinal $W_+W_-$ is
\bdm
  \frac{k_2^{\nu}}{M}\frac{k_1^{\mu}}{M}\M_{\mu \nu}=
  \frac{g}{M} k_1\cdot k_2 = \frac{g}{2 M}(m^2 -i m\Gamma)
       \label{tree5}
\edm 
The width in (\ref{tree5}) is the total width, including decay into
$Z_0 Z_0$. This is 
$$
  \Gamma = \frac{g^2}{64\pi}\frac{m^3}{M^2}\frac{3}{2}
$$
Substiituting this into (\ref{tree5}) gives
\bdm
 \frac{k_2^{\nu}}{M}\frac{k_1^{\mu}}{M}\M_{\mu \nu}= \frac{g}{2}\frac{m^2}{M}
      (1-i\pi\frac{g^2}{16\pi^2}\frac{m^2}{M^2}\frac{3}{8})
     \label{tree6}
\edm
For decay into $\phi_+ \phi_-$
\bdm
   -\M = \frac{g}{2}\frac{m^2}{M}    \label{tree7}
\edm
Again, no imaginary shift is generated because there is no explicit 
dependence on $ K^2 $,

  Renormalization of the truncated Lagrangian is substantially simpler than  
 the renormalization 
of the complete Electroweak \sm \cite{f}. We use the renormalization scheme 
and \rczs from \cite{bw}. The Feynman diagrams to be calculated are just 
those of the first and third columnes of figures 2,3,4, with one addition 
to the first column of fig.3. Because the \sm has both charged and neutral 
Goldstone bosons, there are two additional(equal) triangle diagrams for
 $h\ra W_+W_-$ in which all three internal lines are Goldstone boson 
 propagators. We denote these diagrams 2.f. The evaluation of these 
Feynman diagrams follows from the Feynman rules deduced from the Lagrangian 
(\ref{trunc}). The results are given in Table 2. To evaluate the contributions 
from the counter terms (fig.4), we require the counter terms from \cite{bw}.
Since there is no dynamics from the gauge sector in the Lagrangian 
(\ref{trunc}),$\delta Z_W = 0,\;\;\delta Z_g = 0$.                         Then, $Z_{4 g}= Z Z_W Z_g =Z$.
In \cite{bw},just as in the gaugeless limit of the Abelian Higgs model considered above, also $\delta\zeta_v=0 $ 
and $Z = Z_{\phi}(=Z_{\phi}
^{\star})$. So the counter terms are ($Z_{4\lambda}= Z^2 Z_{\lambda}$)
\bdm
 \frac{g}{2}\frac{m^2}{M}\delta Z = \frac{g}{2}\frac{m^2}{M}\frac{g^2}
{16\pi^2}\frac{m^2}{M^2}[-\frac{1}{8}]  \label{ct1}
\edm
\bdm 
 \frac{g}{2}\frac{m^2}{M}\delta Z_{4\lambda} =\frac{g}{2}\frac{m^2}{M}
\frac{g^2}{16\pi^2}\frac{m^2}{M^2}[\frac{3}{2}\De+\frac{23}{8}-                   \frac{3\sqrt{3}}{8}\pi]
    \label{ct2}
\edm
which appear as the last entries in Table2.  

\begin{table}
\begin{center}

\begin{tabular}{|c|c|c|}
   &
$-i\frac{k_2^{\nu}}{M}\frac{k_1^{\mu}}{M}\underline{\tilde{\tau}}_{\mu \nu}$ &
$ i\underline{\tilde{\tau}}$  \\  [.1in]\hline 
 a   &$ -\frac{3}{8}\Delta_{\epsilon} -\frac{3}{4} +\frac{\sqrt{3}}{8}\pi$ &   
$ -\frac{3}{8}\Delta_{\epsilon} -\frac{3}{4} +\frac{\sqrt{3}}{8}\pi$ \\
 b  &$ -\frac{3}{8}\Delta_{\epsilon} -\frac{3}{4} +i\pi(-\frac{3}{8})$  &   
$ -\frac{5}{8}\Delta_{\epsilon} -\frac{5}{4} +i\pi(-\frac{5}{8})$  \\
 c  &     &$ 2[-\frac{1}{4}\De -\frac{1}{4}]$  \\
 d  & $\frac{3}{8}\De +\frac{3}{2} -\frac{3\sqrt{3}}{8}\pi+\frac{\pi^2}{12}$ &
    $\frac{\pi^2}{12}$ \\
 e  &$ \frac{1}{8}\De +\frac{\pi^2}{48}+i\pi(\frac{1}{4}\ln{2}-\frac{1}{8})$ & 
  $\frac{\pi^2}{48}+i\pi(\frac{1}{4}\ln{2})$  \\ 
 f &  $2[\frac{1}{8}\De +\frac{1}{4}+i\pi(\frac{1}{8})]$  & \\
 ct  & $-\frac{1}{8}$ &  $\frac{3}{2}\De +\frac{23}{8}-\frac{3\sqrt{3}}{8}
    \pi$ 
\end{tabular}
\caption{The one-loop vertex contributions in the \sm. 
Each entry is multiplied by
  common factor  $\frac{1}{2}\frac{g^3}{16\pi^2}\frac{m^4}{M^3}$} 
\end{center}
\end{table}

For the external line factors there is no contribution 
from $Z=Z^{\star}$, but we need
\bdm
 \delta Z_h^{\star} = \frac{g^2}{16 \pi^2}\frac{m^2}{M^2}[\frac{3}{2}- 
   \frac{\sqrt{3}}{4}\pi]   \label{zh}
\edm
So the external line factor for both processes is
\bdm
   \sqrt{\frac{Z_h^{\star}}{Z}}=1 +\frac{1}{2}(\delta Z_h^{\star}-\delta Z)=
    1+\frac{g^2}{16 \pi^2}\frac{m^2}{M^2}[\frac{13}{16}-\frac{\sqrt{3}}{8}\pi]
    \label{rzh}
\edm

  Combining all of these,including the imaginary part from (\ref{tree6})      gives the result

\bdm
  \frac{k_2^{\nu}k_1^{\mu}}{M^2}\M_{\mu\nu}     =\frac{g}{2}\frac{m^2}{M}       
\{1+\frac{g^2}{16 \pi^2}\frac{m^2}{M^2}[\frac{19}{16}-\frac{3\sqrt{3}}{8}\pi
      +\frac{5}{48}\pi^2 +\frac{i\pi}{4}(\ln{2}-\frac{5}{2})]\}   
 = -\M
\label{rsm}
\edm

The real part agrees with old one-loop 
calculations \cite{fj} of both amplitudes .
 Recent two-loop calculations \cite{gh}
based on the \ET have obtained the imaginary part of $ -\M$ in agreement
with (\ref{rsm}).

\section{\bf Discussion}
    
   The working out of the \ET in this example is rather remarkable. It 
provides further testimony to the importance of the complex pole mass of 
the unstable particle (see also {\cite{s ps}). This author's original thought 
was that treatment of the external Higgs boson mass could not affect the 
\ET because it would be common to all the processes related by the \ET ;       
but in some cases $m^2$  appearing in the tree amplitude is a  
renormalized Lagangian parameter with no imaginary part,
and in other cases it is the square of the Higgs boson four-momentum 
which does have an imaginary part. 

   The calculation is also consistent with the general principle of 
renormalized perturbation theory that analyticity and unitarity 
require real renormalization constants. Renormalization only provides 
subtractions for the real parts of self-energy functions. A problem 
which will have to be faced in higher order is the different real    
renormalized masses $m_r^2$ and $m_p^2$ discussed below (\ref{Dh})
 \cite{s ps}. 

\appendix
\section{\bf Renormalization constants and Vev in the Abelian Higgs model}

The renormalization constants and vev which satisfy the five specified 
conditions are calculated in the leading order $g^2\frac{m^2}{M^2}$, which 
remains finite in the limit considered. (It is convenient to calculate the 
scalar vertex \rcz in place of $Z_{\lambda}$)

$$
  \delta\zeta_v = 0
$$
$$
  \delta Z_A = 0
$$
$$
  \delta Z = \fac [-\frac{1}{2}]
$$
$$
   \delta Z_{\mu^2} = \fac [2 \Delta_{\epsilon} +7 - \frac{3\sqrt{3}}{2}\pi]
$$
$$
   \delta Z_{4\lambda} = \fac [5\Delta_{\epsilon} + \frac{19}{2} 
                                     -\frac{3\sqrt{3}}{2}\pi]
$$ 
In these formulas $\Delta_{\epsilon}$ is the usual dimensional regularization 
quantity $\frac{2}{4-d}-\gamma_E + \ln{4\pi}$ .  

For use in (\ref{rzh}), we also need the star-Z's
$$
  \delta Z^{\star}_h = \fac [5 -\sqrt{3}\pi]
$$
$$
  \delta Z^{\star}_{\chi} = \fac [-\frac{1}{2}]
$$

\section{Acknowledgements}

I particularly thank an anonymous referee for his/her hunch that the \ET
shuld work out if the decay amplitude of the unstable Higgs boson is 
evaluated at its complex on-shell value.

I thank Y.-Y.Charng for checking some of the calculations.

\end{document}